%% file: xpm.tex
\documentclass{article}
\usepackage{fullpage}
\usepackage{authblk}
\usepackage[numbers]{natbib}
\usepackage{nicefrac}
\usepackage{pgfplots}
\pgfplotsset{compat=1.18}
\usetikzlibrary{patterns, fit, backgrounds}

\colorlet{warn}{red!80!black}
\usepackage{multirow}
\usepackage{amsmath}
\usepackage{amssymb}
\usepackage{graphicx}
\usepackage[colorlinks=true]{hyperref}
\usepackage[font=small,labelfont=bf]{caption}
\providecommand{\keywords}[1]{\textbf{\textit{Keywords---}} #1}

\begin{document}

\title{Enhanced Permeability Estimation in Microporous Rocks Using a Hybrid Macropore-Darcy Approach}

\author[1]{Dmytro Petrovskyy}
\author[2,*]{Julien Maes}
\author[2]{Hannah P. Menke}
\author[2]{Muhammad Ali}
\author[2,3]{Abdul H. Mazeli}
\author[3]{Muhammad Z. Kashim}
\author[3]{Zainol A. A. Bakar}
\author[2]{Kamaljit Singh}


\affil[1]{Independent Consultant,Ivano-Frankivsk,76000,Ukraine}

\affil[2]{Institute of GeoEnergy Engineering,Heriot-Watt University, Edinburgh, EH14 4AS,United Kingdom}

\affil[3]{PETRONAS Research Sdn Bhd, Bangi, Selangor,Malaysia}
\affil[*]{Corresponding author, j.maes@hw.ac.uk}

\maketitle

\begin{abstract}
This study presents a novel workflow for constructing hybrid macropore-Darcy models from micro-CT images of microporous rocks. In our approach, macropore networks are extracted using established methods, while the microporosity is characterised through segmented phase classification and incorporated into the model as Darcy cells. Effectively, Darcy cells capture the micro scale connectivity variations that are missing in the macroscopic networks. This dual entity model thus incorporates both the conventional macroscopic pore structure and the critical flow pathways present in the under-resolved microporous regions. The proposed workflow is rigorously validated by comparing the permeability estimates with direct numerical simulation (DNS) results and experimental measurements. Our findings demonstrate that this hybrid approach reliably reproduces fluid flow behaviour in complex porous media while significantly reducing computational demands, offering a promising tool for advanced groundwater modelling and water resource management.
\end{abstract}

\keywords{pore scale modelling, permeability, microporosity}

\input{body}

\bibliographystyle{spphys} 
\bibliography{cas-refs}

\end{document}

%% file: body.tex
\newcommand{\abr}[1]{{\small#1}}

\input{sec1-intro}
\input{sec2-method}
\input{sec3-validation}

\section{Conclusions}
In this paper, we demonstrated the application of the hybrid macropore-Darcy network approach in models of carbonate rocks that have dual porosity systems. The predicted permeability values well correlate with the corresponding values obtained in more physically-informed and resource demanding \abr{DNS} methods and experiments. Additionally, we discussed the normalisation of the extracted macroscopic networks, with the help of \abr{DNS} on smaller models, to correctly estimate conductances of the corresponding network elements.

We compared results of numerical simulations and experiments on two carbonate samples. We confirmed that the model is able to reproduce \abr{DNS} simulation results for both permeability and pressure field on subsamples of smaller size (500\texttimes1000\texttimes1000 voxels), where each \abr{DNS} simulation is feasible. We then showed that \abr{XPM} is capable of reproducing experimental measurement of permeability on the full cores with good accuracy.

The main advantages of the hybrid network method are its computational runtime efficiency and reduced memory requirements, which are a result of the underlying linear system flow assumptions. Therefore, not only does it become faster to obtain modelling results, but we can also host larger models on existing hardware infrastructure.

Finally, the numerical implementation is open source and publicly available as \abr{XPM} software that facilitates further contributions and integration in other systems. Apart from the modelling toolset, \abr{XPM} streamlines rendering in \abr{3D} without the necessity for additional visualisation software.

\section*{Acknowledgements}
D.P., J.M., H.P.M. and K.S acknowledge funding from PETRONAS. H.P.M and K.S. acknowledge funding from the ECO-AI project, EPSRC grant reference EP/Y006143/1.

%% file: sec1-intro.tex
\section{Introduction}

Subsurface porous media, such as groundwater aquifers, hydrocarbon reservoirs, geothermal sites \cite{Crooijmans2016} or greenhouse gas storage units \cite{Onoja2018, Reynolds2018}, commonly exhibit multiple scales of geological heterogeneity \cite{Menke2022}.  It is vital to understand the underlying pore structure, morphology and physics for accurate evaluation of sub surface reservoir performance. Representative and reliable numerical modelling of such natural formations requires the application of sophisticated multiscale approaches as traditional single scale methods often cannot capture the necessary level of complexity \cite{Bultreys2015, Carrillo2019, Prodanovic2015}.

Modern computed tomography (\abr{CT}) equipment can readily produce digital pore scale images of geological rocks in varying resolutions \cite{Cnudde2013, Knackstedt2009}. Images of higher resolution better reveal smaller scale features that may be crucial in governing fluid flow. However, the improved resolution also limits the overall imaging extent and frequently does not capture the representative elementary volume essential for the acquisition of characteristic rock properties. Consistent and coherent integration of data from multiple scales of heterogeneity remains a substantial challenge for successful predictive modelling in such porous systems.

Three distinct classes of numerical methods are capable of pore scale fluid flow simulation -- Lattice-Boltzmann methods (\abr{LBM}), direct numerical simulations (\abr{DNS}) and pore network modelling (\abr{PNM}). \abr{LBM} and \abr{DNS} are more physically informed and computationally demanding as they solve the flow equations directly on three-dimensional (\abr{3D}) images \cite{Maes2018, Pan2004, Raeini2014}. On the other hand, \abr{PNM} is a simplified pore scale model that preserves characteristic geometrical and topological features, such as pore size distribution and connectivity patterns, and acts as an effective coarsening of the underlying pore space \cite{Jiang2007, Raeini2017}. Fluid modelling in \abr{PNM} is approximated by predetermined displacement rules and simplified flow equations \cite{Petrovskyy2021, Valvatne2004}.

An extension to \abr{DNS} methods for multiscale modelling is the Brinkman formulation that marries the Stoke's and Darcy's terms in a single expression \cite{Carrillo2019, Krotkiewski2011}. Similarly to the single scale \abr{DNS}, the Brinkman approach remains a numerically expensive task for large models. The integration of under-resolved porous regions in \abr{PNM} can be achieved by the explicit introduction of large stochastically generated subnetworks \cite{Jiang2013} or the addition of the upscaled microlinks \cite{Bultreys2015, Foroughi2024}. The primary challenge to both is the systematic and robust allocation of the corresponding auxiliary entities within the original macroscopic network. The difficulty becomes particularly prominent when the under-resolved porosity is non-homogeneous, i.e. the distribution of greyscale values in a digital image has multiple peaks or does not exhibit them altogether. Furthermore, the microlinks are constrained to only directly enhance the throughput between the existing pairs of macroscopic pores, which does not modify the size, specifically the row count, of the resultant pressure matrix reflecting the modelled porous medium.

In our work, we aim to address the shortcomings of the discussed multiscale \abr{PNM} methods by applying an alternative set of entities referred to as Darcy cells that complement existing macroscopic pore network nodes and throats \citep{Shi2024, Zhang2024}. Physically, the Darcy cells correspond to under-resolved regions that are characterized by its porosity and permeability. While the Darcy cells and microlinks exhibit similarities in terms of flow upscaling, their topological roles differ. The Darcy cells function as vertices within the connectivity graph, whereas microlinks serve as edges. The proposed model is implemented in the \abr{XPM} (Extensive Pore Modelling) software that integrates and facilitates the creation, simulation and volumetric visualisation of the hybrid pore network models. Finally, our development is open source that is freely and readily available to the wider audience at \href{https://github.com/dp-69/xpm}{https://github.com/dp-69/xpm}.

The next section presents the methodology for hybrid network construction and flow calculation. Then, we outline two microporous samples to validate the modelling results. Finally, the \abr{XPM} results are compared and verified against equivalent \abr{DNS} simulations and experimental results.

%% file: sec2-method.tex
\section{Methodology}

\subsection{Hybrid network construction}

\begin{figure*}[!t]
\centering
\begin{tikzpicture}
\color{black} 
\node[inner sep=0pt, outer sep=0pt] (slice) at (0, 0) {
  \includegraphics[width=0.6\textwidth]{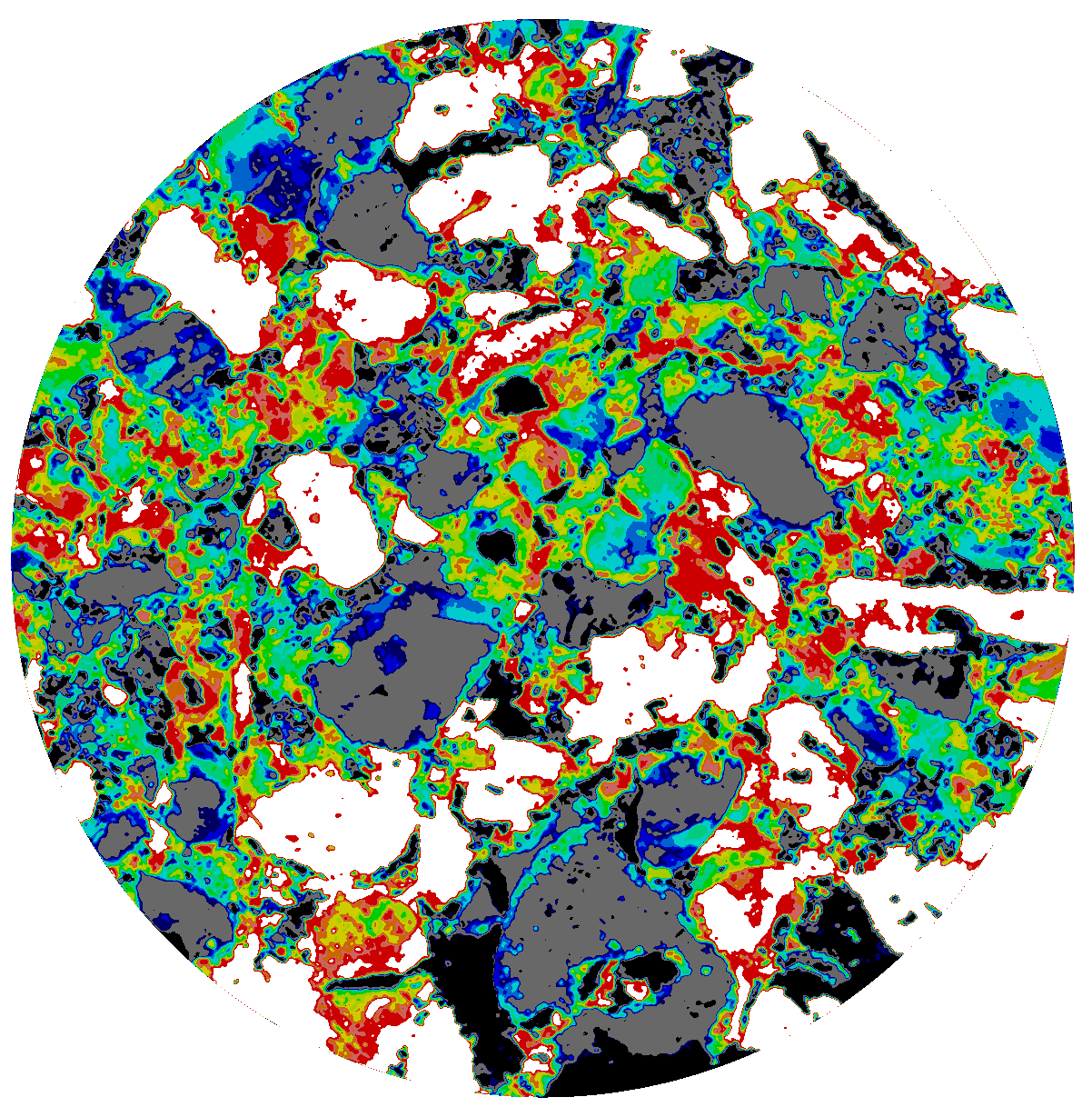}};
\begin{scope}[shift={(current bounding box.east)}, xshift=0.03\textwidth, yshift=32.5mm, overlay, inner sep=0pt, outer sep=0pt]

  \begin{scope}[every node/.style={draw, very thin, anchor=west, minimum width=3mm, minimum height=3mm}]
    \node[fill=black] at (0, 0) {};
    \node[fill={rgb,255 : red,100; green,100; blue,100}] at (0, -5mm)  {};
    \node[fill={rgb,255 : red,0;   green,0;   blue,100}] at (0, -10mm) {};
    \node[fill={rgb,255 : red,0;   green,0;   blue,204}] at (0, -15mm) {};
    \node[fill={rgb,255 : red,0;   green,100; blue,204}] at (0, -20mm) {};
    \node[fill={rgb,255 : red,0;   green,204; blue,204}] at (0, -25mm) {};
    \node[fill={rgb,255 : red,0;   green,204; blue,100}] at (0, -30mm) {};
    \node[fill={rgb,255 : red,0;   green,204; blue,0  }] at (0, -35mm) {};
    \node[fill={rgb,255 : red,174; green,204; blue,0  }] at (0, -40mm) {};
    \node[fill={rgb,255 : red,204; green,204; blue,0  }] at (0, -45mm) {};
    \node[fill={rgb,255 : red,204; green,100; blue,0  }] at (0, -50mm) {};
    \node[fill={rgb,255 : red,204; green,100; blue,100}] at (0, -55mm) {};
    \node[fill={rgb,255 : red,204; green,0;   blue,0  }] at (0, -60mm) {};
    \node[fill=white] at (0, -65mm) {};
  \end{scope}

  \begin{scope}[every node/.style={anchor=west, xshift=4mm}]
    \node[] at (0, 0)     {Pore};
    \node[] at (0, -5mm)  {1};
    \node[] at (0, -10mm) {2};
    \node[] at (0, -15mm) {3};
    \node[] at (0, -20mm) {4};
    \node[] at (0, -25mm) {5};
    \node[] at (0, -30mm) {6};
    \node[] at (0, -35mm) {7};
    \node[] at (0, -40mm) {8};
    \node[] at (0, -45mm) {9};
    \node[] at (0, -50mm) {10};
    \node[] at (0, -55mm) {11};
    \node[] at (0, -60mm) {12};
    \node[] at (0, -65mm) {Solid};
  \end{scope}
\end{scope}
\begin{scope}[shift={(current bounding box.south west)},
  xshift=-0.06\textwidth, yshift=0.040\textwidth, overlay]
  \draw (0, 0) -- (0.12581\textwidth, 0) node[midway, above] {1 mm};
  \draw (0, 0.75ex) -- (0, -0.75ex);
  \draw (0.12581\textwidth, 0.75ex) -- (0.12581\textwidth, -0.75ex);
\end{scope}
\end{tikzpicture}
\caption{A slice extracted from the center of an Estaillades micro-CT image (6000\texttimes1202\texttimes1218 voxels with 3.9676 $\mu$m resolution) containing 14 labels.
Black and white labels correspond to the resolved pore and solid states, respectively, while the remaining colours are associated with the under-resolved regions.
}
\label{fig:EstailladesSlice}
\end{figure*}
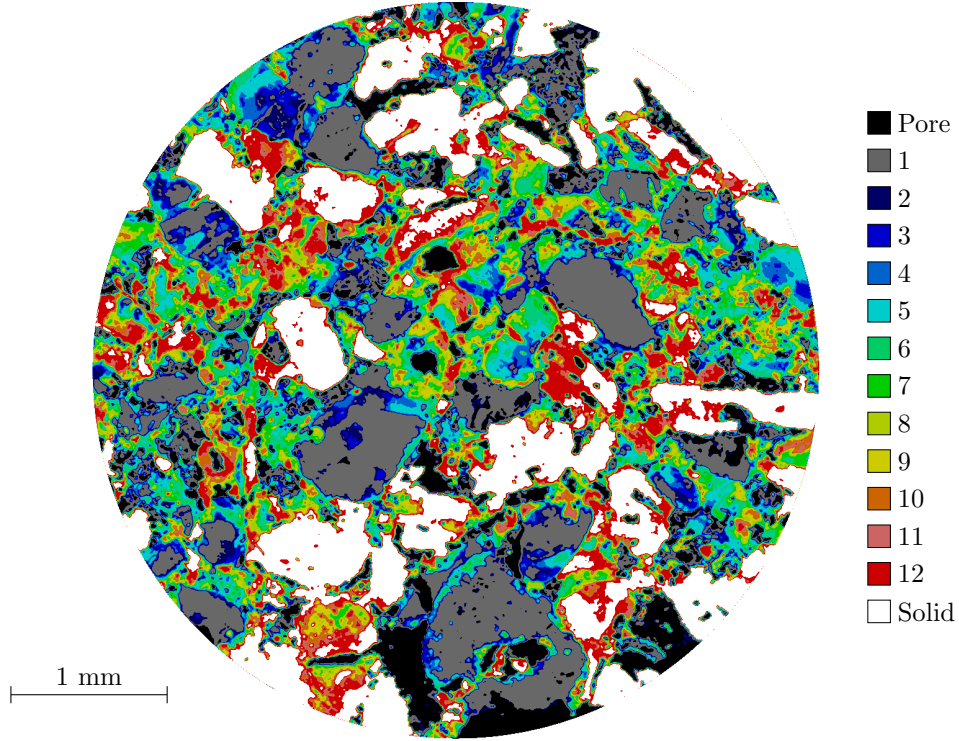

The starting point of our modelling workflow is a \abr{3D} image that has been segmented into a set of labels, which includes at least two labels for the resolved pore and solid voxels, as well as several labels for the intermediate under-resolved voxels. An example of such an image is presented in Figure~\ref{fig:EstailladesSlice}, which shows a single slice from a large micro-\abr{CT} image of Estaillades carbonate \citep{Menke2022}. The porosity and permeability values across under-resolved voxels of the same label are assumed to be equal for the purpose of modelling.

\begin{figure*}[!t]
\centering
\begin{tikzpicture}[color=black]
  \node[anchor=south west, align=center, inner sep=0pt, outer sep=0pt] (tl) at (0,              0) {
    \includegraphics[width=0.35\textwidth]{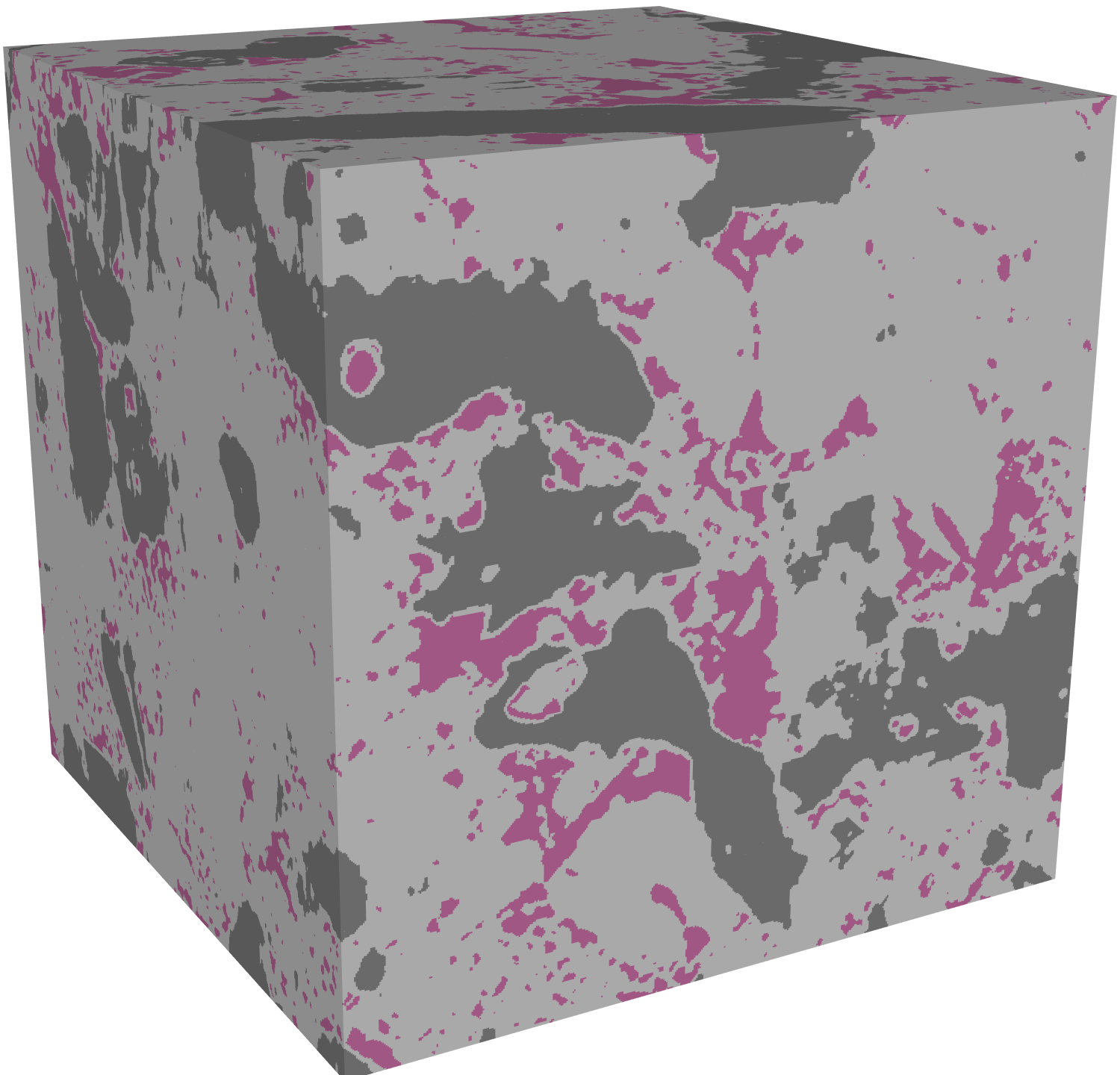}\\\footnotesize (a) Input image};
  
  \node[anchor=south west, align=center, inner sep=0pt, outer sep=0pt] (tr) at (0.38\textwidth, 0) {
    \includegraphics[width=0.35\textwidth]{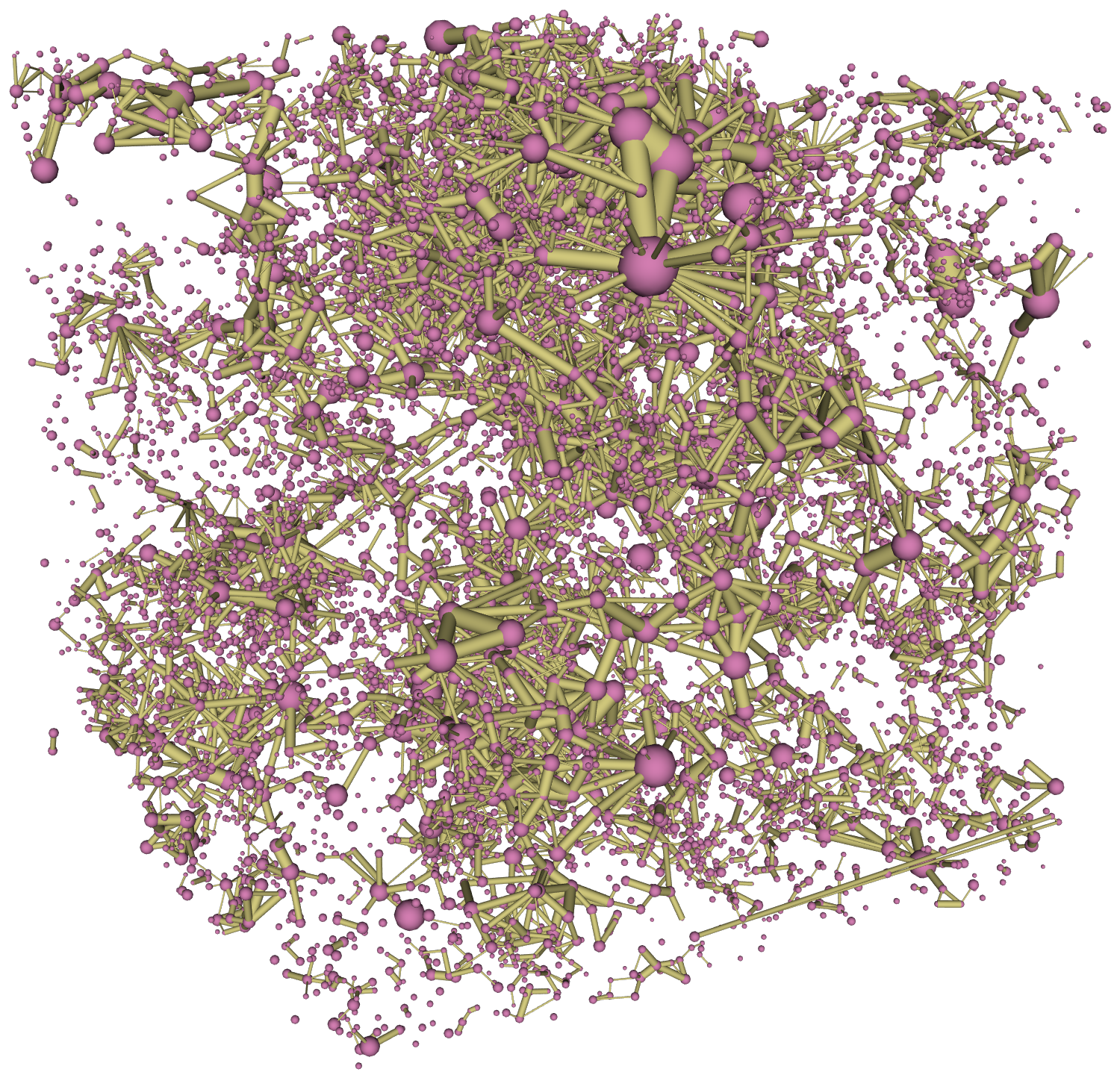}\\\footnotesize (b) Extracted macro network};
  
  \node[anchor=north west, align=center, inner sep=0pt, outer sep=0pt] at (0, -1em) {
    \includegraphics[width=0.35\textwidth]{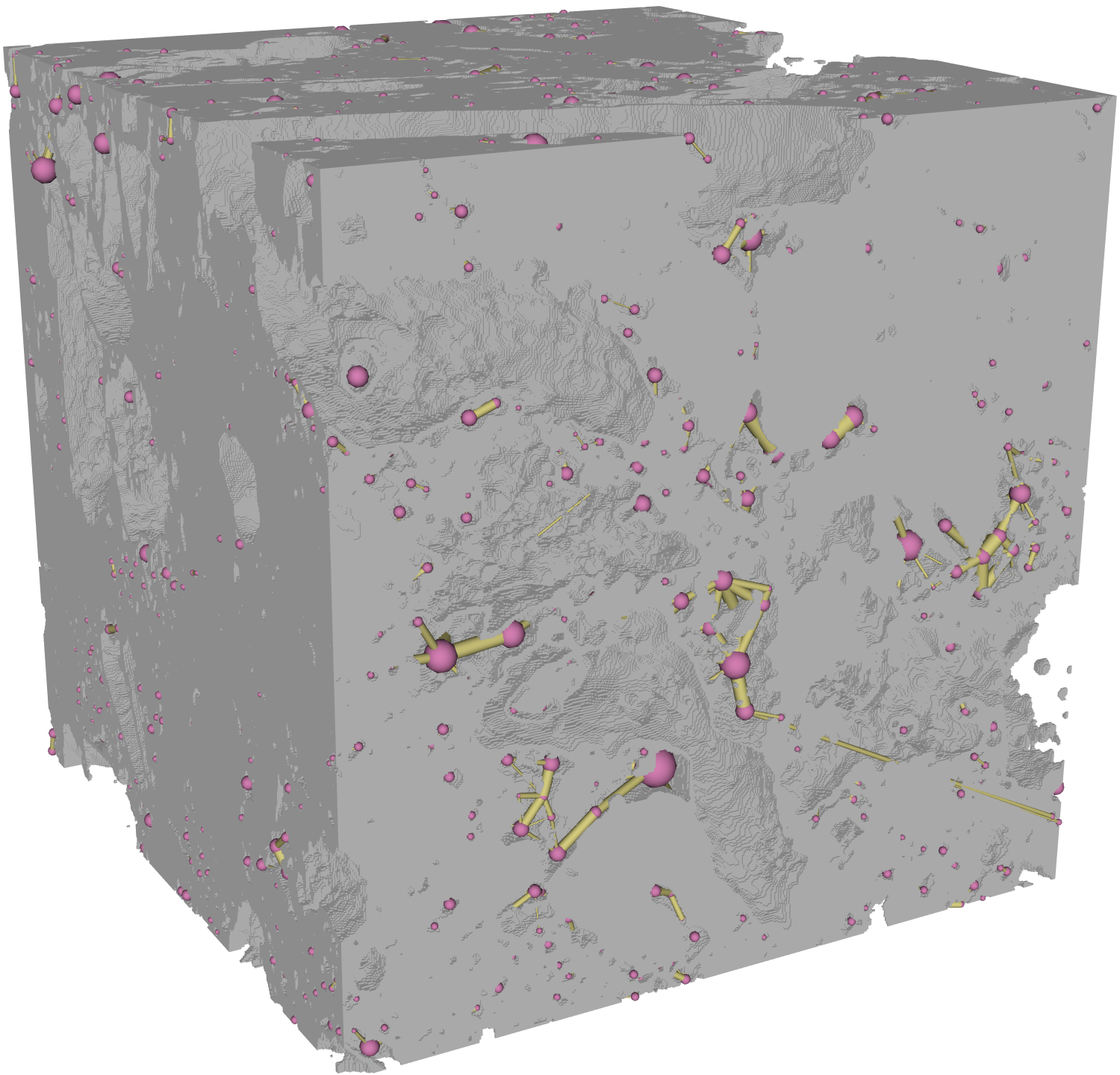}\\\footnotesize (c) Hybrid network};
  
  \node[anchor=north west, align=center, inner sep=0pt, outer sep=0pt] at (0.38\textwidth, -1em) {
    \includegraphics[width=0.35\textwidth]{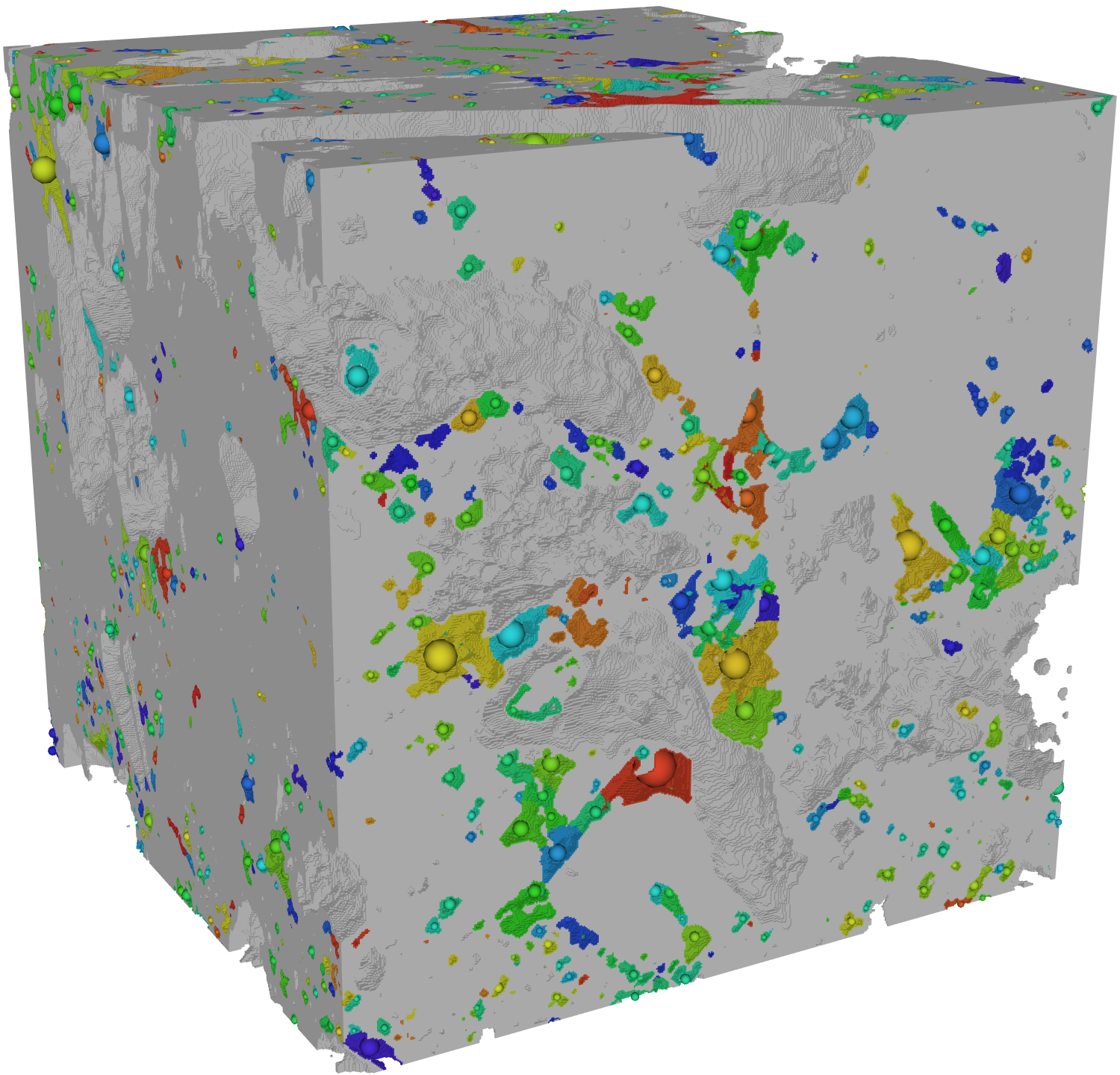}\\\footnotesize (d) Cross-entity relationship};

  \begin{scope}[shift={(tl.north west)}]
    \draw[<->] (0.000\textwidth, -0.004\textwidth) -- (0.203 \textwidth, 0.008\textwidth) node[midway, above, sloped] {\footnotesize 1.98 mm};
  \end{scope}
\end{tikzpicture}


\caption{A 3D volumetric image displaying pore (purple), under-resolved (light grey), and solid (dark grey) voxels of a 500\textsuperscript3 Estaillades carbonate subsample (a). The extracted macro nodes and throats are shown as purple spheres and yellow cylinders, respectively (b, c). The cross-entity connection between the under-resolved voxels and the corresponding macro nodes is emphasised by the same colour (d). }
\label{fig:network-extract}
\end{figure*}

The first step is the extraction of the macroscopic network from the resolved pore voxels. The resultant network consists of the macroscopic nodes (macro nodes) connected by throats. Figure \ref{fig:network-extract} shows a \abr{3D} rendering of an Estaillades subsample and the corresponding macroscopic network acquired using \emph{pnextract}, an open-source software \citep{Raeini2017}. The second step is network enhancement using the under-resolved information, or more specifically, continuum entities referred to as Darcy cells.

A Darcy cell is a finite volumetric cell of constant porosity and permeability. There is a strong similarity between a conventional reservoir modelling cell and a Darcy cell except for their scale of application. In general, Darcy cells can be of any shape, but in our case they match each under-resolved voxel of a \abr{CT} image. The connectivity between Darcy cells is straightforward to construct by detecting all pairs of the adjacent under-resolved voxels, with at most six connections per voxel. Finally, the interconnection between the macroscopic network and the Darcy cells is facilitated by a network extraction attribute \textsc{velem}, that uniquely relates each resolved pore voxel to a single macro node. Therefore, pairs of adjacent under-resolved and resolved pore voxels, together with \textsc{velem}, establish the missing connection between Darcy cells and macro nodes in the ultimate hybrid network, as shown in Figure~\ref{fig:network-extract}.

\subsection{Flow calculation}

The hybrid network flow model comprises three distinct mechanisms. 
\begin{enumerate}

\item A conventional macroscopic flow between two macro nodes $i$ and $j$ separated by a throat $t$, as described by Hagen-Poiseuille equation
\begin{subequations}
\label{eq:m}
\begin{align}  
  & q_{ij} = -\gamma_{ij} \Delta p \\
  & \gamma_{ij}^{-1} = \frac{L_i}{g_i} + \frac{L_t}{g_t} + \frac{L_j}{g_j} \\
  \label{eq:hagen-poise} & g = h \frac{A^2 G}{\mu}
\end{align}
\end{subequations}
where $q$ is the flow rate, $g$ is the conductance, $L$ is the characteristic length, $h$ is a multiplier, $A$ is the cross-section area, $G$ is the shape factor, and $\mu$ is the viscosity.

\item A standard flow in a porous medium across the interface of two neighbouring Darcy cells $c$ and $d$, as defined by Darcy's law
\begin{subequations}
\label{eq:d}
  \begin{align}
  \label{eq:darcy-law} & q_{cd} = -\frac{k A}{\mu L_d} \Delta p \\
  & k = 2 \left( \frac1{k_{c}} + \frac1{k_{d}} \right)^{-1}
\end{align}
\end{subequations}
where $k$ is the permeability.
  
\item A cross-entity flow between macro nodes and Darcy cells, expressed as
\begin{subequations}
\label{eq:md}
\begin{align}
  & q_{id} = -\gamma_{id} \Delta p \\  
  & \gamma_{id}^{-1} = \frac{L_i}{g_i} + \frac{L_t}{g_i} + \frac12\frac{\mu}{k L_d}
\end{align}
\end{subequations}

\end{enumerate}

Macroscopic properties \eqref{eq:m} are obtained directly from the extracted pore network model. Darcy properties \eqref{eq:d} are available for all neighbouring pairs of under-resolved voxels, as long as permeability is defined for each voxel (e.g., by defining permeability for each label). In our model, we assume that all macro nodes and throats have equilateral triangle shapes with the following properties
\begin{equation}
  \label{eq:mult}
  h = \frac34 \lambda \qquad G = \frac{\sqrt{3}}{36}
\end{equation}
where $\lambda$ is a multiplying factor. These pore shape assumptions (with $\lambda=1$) yield accurate results for sandstones \citep{Bakke97, Oren98}, but underestimate conductances in more complex and heterogeneous pore systems such as the microporous carbonates and require additional correction. A possible solution is to employ other shapes, such as polygons or stars \citep{Patzek01, Ryazanov09}. In the current model, instead of fitting the shapes of nodes and throats, we normalise the macroscopic conductances to match the permeability from \abr{DNS} by fitting the dimensionless multiplier ($\lambda>1$).

The discussed hybrid network model is implemented in an open source software \abr{XPM}. The software is based on the C++ programming language, which considerably optimises resource consumption and maximises runtime performance. The implementation employs the parallel algebraic multigrid method \emph{BoomerAMG}~\citep{Henson02} to solve the pressure linear system. The method uses the commonly available Message Passing Interface (\abr{MPI}) computing architecture to parallelise the underlying algorithm. \emph{BoomerAMG} is an unopinionated tool that offers the highest degree of setup granularity and potential speed-up but requires extensive explicit configuration, most notably the problem decomposition across parallel \abr{MPI} processes.

Geometrically, our algorithm compartmentalises the modelling domain into evenly spaced volumetric blocks. Each block hosts the macro nodes and Darcy cells, the centres of which are located inside the block. Algorithmically, each \abr{MPI} process handles a dedicated block and owns the corresponding set of rows within the parallelised framework. Naturally, the block count is equal to the \abr{MPI} process count launch argument. For optimal hardware utilisation, the advice is to configure \abr{XPM} input such that the block count is less than or equal to the available processor core count.

%% file: sec3-validation.tex
\pgfdeclareverticalshading{shadebot}{50bp} {
  rgb(0bp)=(0.231373, 0.298039, 0.752941);
  rgb(30bp)=(0.548188, 0.581521, 0.808972);
  rgb(50bp)=(0.865003, 0.865003, 0.865003)  
}

\pgfdeclareverticalshading{shadetop}{50bp} {  
  rgb(0bp)=(0.865003, 0.865003, 0.865003);
  rgb(30bp)=(0.785443, 0.440345, 0.507012);
  rgb(50bp)=(0.705882, 0.0156863, 0.14902)
}

\newcommand{\pressurelegendbar}{
  \begin{scope}[shift={(current bounding box.east)}, xshift=0.04\textwidth, overlay]
  \node[
    rectangle,
    shading=shadetop,
    anchor=south west,
    draw=none,
    minimum width=2mm,
    minimum height=3cm,
    inner sep=0pt,
    outer sep=0pt,
  ] (bartop) {};
  
  \node[
    rectangle,
    shading=shadebot,
    anchor=north west,
    draw=none,
    minimum width=2mm,
    minimum height=3cm,
    inner sep=0pt,
    outer sep=0pt
  ] (barbot) {};
  
  \node[anchor=west] at (barbot.south east) {0};
  \node[anchor=west] at (bartop.south east) {0.005};
  \node[anchor=west] at (bartop.north east) {0.01};
  \node[anchor=south west, shift=({-0.2ex, 0.75em}), inner sep=0pt, outer sep=0pt] at (bartop.north) {P (kPa)};
  
  \node[
    anchor=north west,
    fill=black,
    draw=none,
    rectangle,
    minimum width=0.8mm,
    minimum height=0.2pt,
    inner sep=0pt,
    outer sep=0pt] at (bartop.north east) {};
  \node[
    anchor=west,
    fill=black,
    draw=none,
    rectangle,
    minimum width=0.8mm,
    minimum height=0.2pt,
    inner sep=0pt,
    outer sep=0pt] at (bartop.south east) {};
  \node[
    anchor=south west,
    fill=black,
    draw=none,
    rectangle,
    minimum width=0.8mm,
    minimum height=0.2pt,
    inner sep=0pt,
    outer sep=0pt] at (barbot.south east) {};
  \end{scope}
}

\section{Validation}

\subsection{Samples}
Two microporous samples are considered: a PETRONAS sample acquired from Central Luconia (referred to as \abr{MCR}) and a sample of Estaillades carbonate rock. Although both samples are microporous, they have different structures. \abr{MCR} has macro-pores with a minimum size of 5 $\mu$m and a mean size of 12 $\mu$m that are poorly connected among themselves, but well connected through a wide range of micro-pores, with size varying from a few nanometres (nm) to a few $\mu$m. The Estaillades sample has a bimodal pore size distribution with macropores of average size 6 $\mu$m and micropores of average size 30 nm, as described in \citep{Menke2022}. The macropores are mostly connected. Both samples are imaged using X-ray micro-CT imaging, with setup summarised in Table~\ref{Tab:CT}. For Estaillades, the sample was imaged dry and doped with 25 weight percent potassium iodide brine. The images were registered and subtracted to obtain a differential image showing the amount of pore space connected as the pore space occupied by the doped brine. This connected pore space image was then segmented into pore (fully resolved), 12 intervals of microporosity increasing in connected porosity, and solid (no porosity or unconnected). This 14 interval segmentation was then used inside our \abr{DNS} and \abr{XPM} simulators for permeability calculation. For \abr{MCR}, a simplified approach was used where the fully resolved pores were segmented with watershed, and then the rest of the samples was split into 4 microporous intervals and solid using a user determined visual threshold.  

\begin{table}
\caption{Parameters of CT scan imaging for MCR and Estaillades samples.}
\renewcommand{\arraystretch}{1.5}
\centering
\begin{tabular}{cccc}
\hline
Sample & Resolution ($\mu$m) & Size \\
\hline
MCR & 5.5 & 3000\texttimes1500\texttimes1500 \\
Estaillades & 3.9676 & 6000\texttimes1202\texttimes1218 \\
\hline
\end{tabular}
\label{Tab:CT}
\end{table}

In order to estimate the permeability of each microporous label in \abr{MCR}, the results from \cite{WenPin2024} were used. In this study, an analogue sample was imaged at a resolution of 0.65 $\mu$m using using the Tomographic Microscopy Coherent rAdiology experimenTs (\abr{TOMCAT}) beamline at Swiss Light Source (\abr{SLS}) in Villingen Switzerland. The results were used to obtain a porosity-permeability relationship which is applied to the microporous intervals. In \cite{WenPin2024}, three different cases (low, mid and high) were investigated, and here we used the low case. The results are presented in Table \ref{tab:segment_mcr}. For Estaillades, a nano-CT image with 32 nm resolution was obtained from a laser-cut subsample. The information from the image was then used as input into an object-based pore network generator, on which permeability fields were simulated for a range of porosities, creating a synthetic porosity-permeability relationship that was used for each label. The results are also presented in Table \ref{tab:segment_esta}.

\begin{table}
\caption{Properties of segmentation intervals for MCR.}
\renewcommand{\arraystretch}{1.5}
\centering
\begin{tabular}{cccc}
\hline
Label & Fraction & $\phi$ & $k$ (mD) \\
\hline
Pore & 0.12 & 1.0 & -- \\[-1ex]
1 & 0.1 & 0.6 & 90 \\[-1ex]
2 & 0.08 & 0.4 & 15 \\[-1ex]
3 & 0.11 & 0.2 & 0.7 \\[-1ex]
4 & 0.09 & 0.05 & 0.02 \\[-1ex]
Solid & 0.5 & 0 & 0 \\
\hline
\end{tabular}
\label{tab:segment_mcr}
\end{table}

\begin{table}
\caption{Properties of segmentation intervals for Estaillades.}
\renewcommand{\arraystretch}{1.5}
\centering
\begin{tabular}{cccc}
\hline
Label & Fraction & $\phi$ & $k$ (mD) \\
\hline
Pore & 0.100 & 1.0 & -- \\[-1ex]
1 & 0.172 & 0.57 & 7.57 \\[-1ex]
2 & 0.046 & 0.52 & 7.00 \\[-1ex]
3 & 0.046 & 0.47 & 4.85 \\[-1ex]
4 & 0.049 & 0.42 & 3.28 \\[-1ex]
5 & 0.085 & 0.36 & 2.15 \\[-1ex]
6 & 0.072 & 0.27 & 0.817 \\[-1ex]
7 & 0.041 & 0.22 & 0.465 \\[-1ex]
8 & 0.037 & 0.18 & 0.247 \\[-1ex]
9 & 0.032 & 0.15 & 0.119 \\[-1ex]
10 & 0.028 & 0.12 & 0.0502 \\[-1ex]
11 & 0.024 & 0.09 & 0.0175 \\[-1ex]
12 & 0.077 & 0.07 & 0.00482 \\[-1ex]
Solid & 0.192 & 0 & 0 \\
\hline
\end{tabular}
\label{tab:segment_esta}
\end{table}

\subsection{Comparison with DNS}
 In this section, the results of \abr{XPM} are validated by comparison with results obtained using \abr{DNS} with the micro-continuum approach \cite{soulaine2016micro,Maes2024}. In this approach, a computational model that includes all voxels as computational cells is created. Each cell is characterised by its porosity and permeability.  The velocity and pressure field are obtained by solving the Brinkman equation
  \begin{align}
     \nabla\cdot\mathbf{u}=0\\
     -\nabla p + \nabla \cdot \tilde{\mu} \nabla \mathbf{u} - \mu k^{-1}\mathbf{u}=0
 \end{align}
where $\mathbf{u}$ is the fluid velocity, $k$ is the local permeability in the voxel and $\tilde{\mu}$ is the effective viscosity inside the viscosity. The permeability is assumed to be infinite $(k^{-1}=0$) in the macropores and a very small number ($k=10^{-11}$ mD) in the solid. The effective viscosity is an upscaling parameter that takes into account the underlying structure of the velocity field within the unresolved pores inside a voxel. It is a parameter that is difficult to characterise, and often assumed to be a simple function, such as $\mu/\phi$ where $\phi$ is the porosity of the voxel. This model would lead to a no-slip boundary condition at the interface between macropores and micropores. However, \abr{XPM} assumes that the micropores do not affect the conductance between two macropores. This is equivalent to assuming that the tangential velocity at the interface between macropores and micropores is 0. To achieve the same condition in the \abr{DNS} simulation, the effective viscosity inside the micropores is increased, so that%
\begin{equation}
\tilde{\mu}=\frac{100\mu}{\phi}
\end{equation}

To perform this comparison, three subdomains of size 500\texttimes1000\texttimes1000 voxels are extracted from each sample. The simulations are performed on a 64-thread Intel Xeon Gold 6448Y Processor (\abr{CPU}). The \abr{CPU} time is approximately 20 minutes for the \abr{XPM} simulations, and 2 hours for the \abr{DNS} simulations. The \abr{XPM} and \abr{DNS} simulations have similar number of computational cells (approximately 0.3 and 0.2 billion cells for Estaillades and \abr{MCR-1}, respectively), since most of the cells are in the microporous regions which are treated in the same way. Therefore, the \abr{CPU} runtime gain obtained with \abr{XPM} is not strongly related to computational domain sizes. Rather, it is a consequence of the linear nature of the \abr{XPM} equations. The \abr{DNS} simulation solves a non-linear system of equations that is handled with the \abr{SIMPLE} (Semi-Implicit Method for Pressure-Linked Equations) algorithm, to converge the velocity field inside the pores, while for \abr{XPM} this is included in the linear conductance between pores.

For each subsample, a \abr{DNS} simulation that only accounts for the macropores is performed and the permeability $k_\text{ma}$ is calculated. To match the macroscopic permeability obtained by \abr{XPM}, we express the conductance multiplier as follows
\begin{equation}
  \lambda_0=\dfrac{k_\text{macro, DNS}}{k_\text{macro, XPM$_{\displaystyle \lambda\!=\!1}$}}
\end{equation}
\abr{DNS} and \abr{XPM} simulations that account for all segmentation intervals are then performed. The results are presented in Table \ref{Tab:Res}.
 
\begin{table*}[!b]  
\caption{Comparison of DNS and XPM calculated permeabilities for MCR and Estaillades subsamples.}
\renewcommand{\arraystretch}{1.5}
\centering
\begin{tabular}{c cc c ccc}
\hline
\multirow{2}{*}{Sample} & \multicolumn{2}{c}{$k_\text{macro}$ (mD)} & \multirow{2}{*}{$\lambda_0$} & \multicolumn{3}{c}{$k$ (mD)} \\
\cline{2-3}\cline{5-7}
& XPM$_{\lambda=1}$ & DNS & & XPM$_{\lambda=1}$ & DNS & XPM$_{\lambda=\lambda_0}$ \\[0.4ex]
\hline
MCR-1         & 0    & 0    & --   & 1.5  & 1.67 & 1.5 \\
MCR-2         & 3.16 & 9.18 & 2.91 & 13.5 & 18.9 & 22.4 \\
MCR-3         & 32.5 & 79.9 & 2.46 & 44.4 & 90.2 & 93.0 \\
\hline
Estaillades-1 & 3.26 & 6.90 & 2.12 & 11.6 & 17.7 & 16.6 \\
Estaillades-2 & 18.7 & 35.7 & 1.91 & 26.6 & 47.1 & 44.7 \\
Estaillades-3 & 9.7  & 25.8 & 2.66 & 22.2 & 41.5 & 41.4 \\
\hline
\end{tabular}
\label{Tab:Res}
\end{table*}

\begin{figure*}[!t]
\centering

\makeatletter
\tikzset{
        hatch distance/.store in=\hatchdistance,
        hatch distance=5pt,
        hatch thickness/.store in=\hatchthickness,
        hatch thickness=5pt
        }
\pgfdeclarepatternformonly[\hatchdistance,\hatchthickness]{north east hatch}
    {\pgfqpoint{-1pt}{-1pt}}
    {\pgfqpoint{\hatchdistance}{\hatchdistance}}
    {\pgfpoint{\hatchdistance-1pt}{\hatchdistance-1pt}}%
    {
        \pgfsetlinewidth{\hatchthickness}
        \pgfpathmoveto{\pgfqpoint{0pt}{0pt}}
        \pgfpathlineto{\pgfqpoint{\hatchdistance}{\hatchdistance}}
        \pgfusepath{stroke}
    }
\makeatother

\begin{tikzpicture}[trim axis left]
  \pgfdeclarelayer{foreground}
  \pgfdeclarelayer{foreground2}
  \pgfsetlayers{main,foreground,foreground2}

  \begin{axis}[
      color=black,
      ybar,
      ymajorgrids=true,
      grid style={line width=0.1pt, dash pattern=on 3pt off 3pt},
      bar width=6pt,
      width=0.9\linewidth,
      height=8cm,
      enlarge x limits=0.15,
      enlarge y limits=0.05,
      legend cell align={left},
      xticklabels={MCR-1, MCR-2, MCR-3, Estaillades-1, Estaillades-2, Estaillades-3},
      xtick=data,
      ylabel style={rotate=-90, xshift=3ex},
      ylabel={$k$ (mD)},      
      ymin=0,
      ymax=100,
      xtick pos=bottom,
      ytick pos=left,
      ytick={0, 20, 40, 60, 80, 100}, yticklabels={0, 20, 40, 60, 80, 100},
      %
  ]
  \addplot[
    bar width=4pt,  
    pattern=north east hatch,  
    draw=red!75!black,
    pattern color=red!75!black,
    line width=0.2pt,    
    hatch thickness=0.2pt,
    hatch distance=6pt  
  ] plot coordinates {(1, 0) (2, 3.16) (3, 32.5) (4, 3.26) (5, 18.7) (6, 9.7)};

  \addplot[
    bar width=4pt,  
    pattern=north east hatch,  
    draw=blue!75!black,
    pattern color=blue!75!black,
    line width=0.2pt,    
    hatch thickness=0.2pt,
    hatch distance=6pt     
  ] plot coordinates {(1, 0) (2, 9.18) (3, 79.9) (4, 6.9) (5, 35.7) (6, 25.8)};

  \addplot[xshift=3pt, line width=0.2pt, fill=red!50, draw=red!25!black]
    plot coordinates {(1, 1.5) (2, 13.5) (3, 44.4) (4, 11.6) (5, 26.6) (6, 22.2)};
  \addplot[xshift=3pt, line width=0.2pt, fill=blue!50, draw=blue!25!black]
    plot coordinates {(1, 1.67) (2, 18.9) (3, 90.2) (4, 17.7) (5, 47.1) (6, 41.5)};
  \addplot[xshift=3pt, line width=0.2pt, fill=orange!50, draw=orange!25!black]
    plot coordinates {(1, 1.5) (2, 22.4) (3, 93.0) (4, 16.6) (5, 44.7) (6, 41.4)};

\begin{scope}[
  shift={(current bounding box.north east)}, xshift=3ex, yshift=0ex
]
  \begin{pgfonlayer}{foreground2}
    \node[line width=0.2pt, anchor=east, minimum size=8pt, inner sep=0pt, draw=black, pattern=north east hatch, hatch thickness=0.2pt,
    hatch distance=6pt] (l1) at (0, 0) {};
    \node[anchor=west] (r1) at (0, 0) {\vphantom{X}\smash{Macropores}};

    \node[line width=0.2pt, anchor=east, minimum size=8pt, inner sep=0pt, draw=red!25!black, fill=red!50, circle] (l2) at (0, -8) {};
    \node[anchor=west]  at (0, -8) {\vphantom{X}\smash{XPM$_{\lambda=1}$}};

    \node[line width=0.2pt, anchor=east, minimum size=8pt, inner sep=0pt, draw=blue!25!black, fill=blue!50, circle] (l3) at (0, -16) {};
    \node[anchor=west]  at (0, -16) {\vphantom{X}\smash{DNS}};

    \node[line width=0.2pt, anchor=east, minimum size=8pt, inner sep=0pt, draw=orange!25!black, fill=orange!50, circle] (l4) at (0, -24) {};
    \node[anchor=west] (r4) at (0, -24) {\vphantom{X}\smash{XPM$_{\lambda=\lambda_0}$}};
  \end{pgfonlayer}

  \begin{pgfonlayer}{foreground}
    \node[draw=black, fit=(l1)(l2)(l3)(l4)(r1)(r4), inner sep=6pt, fill=white, line width=0.2pt] {};
  \end{pgfonlayer}
\end{scope}

\end{axis}
\end{tikzpicture}

\caption{Comparison of DNS and XPM calculated permeability for subsamples of MCR and Estaillades. For each sample, the first two bars, marked with stripes, correspond to the calculation for macropores only.}
\label{fig:Histo}
\end{figure*}
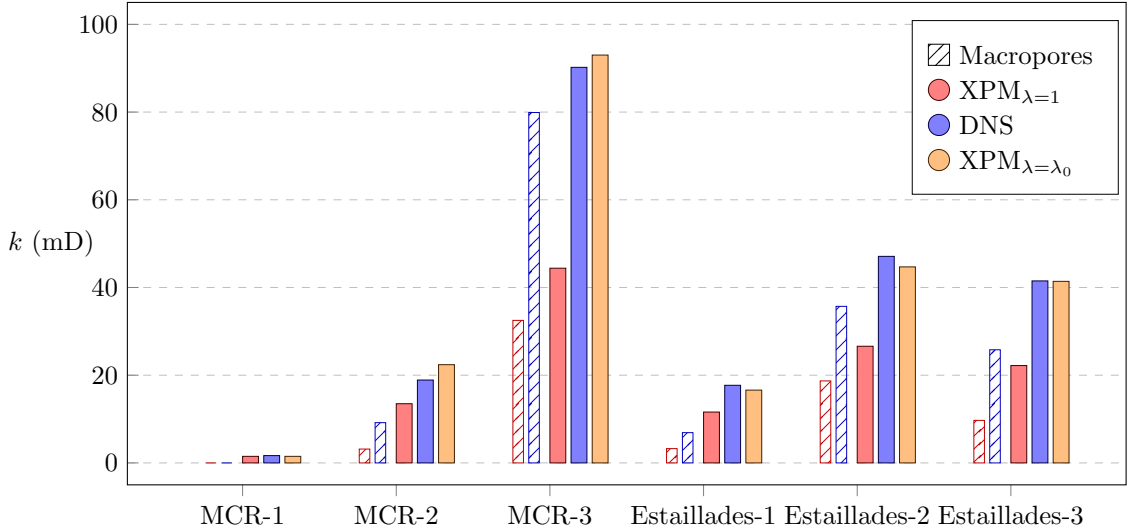

We observe that the permeability calculated with \abr{DNS} is consistently larger than the one calculated with \abr{XPM}. When looking at the difference between \abr{XPM} and \abr{DNS} (Figure~\ref{fig:Histo}), we note that the difference for full permeability is close to the difference for macropores only. Since the microporous regions do not contribute to the flow and are treated as solid in these simulations, we attribute these differences to the calculation of conductance between macropores, which, as previously mentioned, is typically underestimated. The \abr{XPM} macropores simulation is then corrected by using a multiplying factor $\lambda_0$ in the calculation of conductance  for the macroscopic flow equal to the ratio of \abr{DNS} to \abr{XPM} permeability, while the conductance for the under-resolved regions and for the cross-entity flow between macroscopic and under-resolved regions are unchanged. This factor is presented in Table~\ref{Tab:Res}, and is between 1.91 and 2.91 for all subsamples. Thus, the corrected \abr{XPM} simulations give the exact same value than \abr{DNS} for the calculation of the permeability of macropores. The permeability for the full sample is then recalculated and the result are presented in Table~\ref{Tab:Res}. We observe that with this correction, the calculated permeability with \abr{XPM} and \abr{DNS} are close for all samples (Figure \ref{fig:Histo}). The differences between \abr{DNS} and corrected \abr{XPM} simulations are between 0 and 18\% for all samples. \abr{MCR-2} is the sample with the largest difference of 18\% and the one for which the multiplier is the largest that is 2.91. It is the only sample for which the difference is larger than 10\%.

\begin{figure*}[!t]
\centering
\begin{tikzpicture}[color=black]
\node[anchor=south west, align=center, inner sep=0pt, outer sep=0pt] at (0,              0) {
  \includegraphics[width=0.33\textwidth]{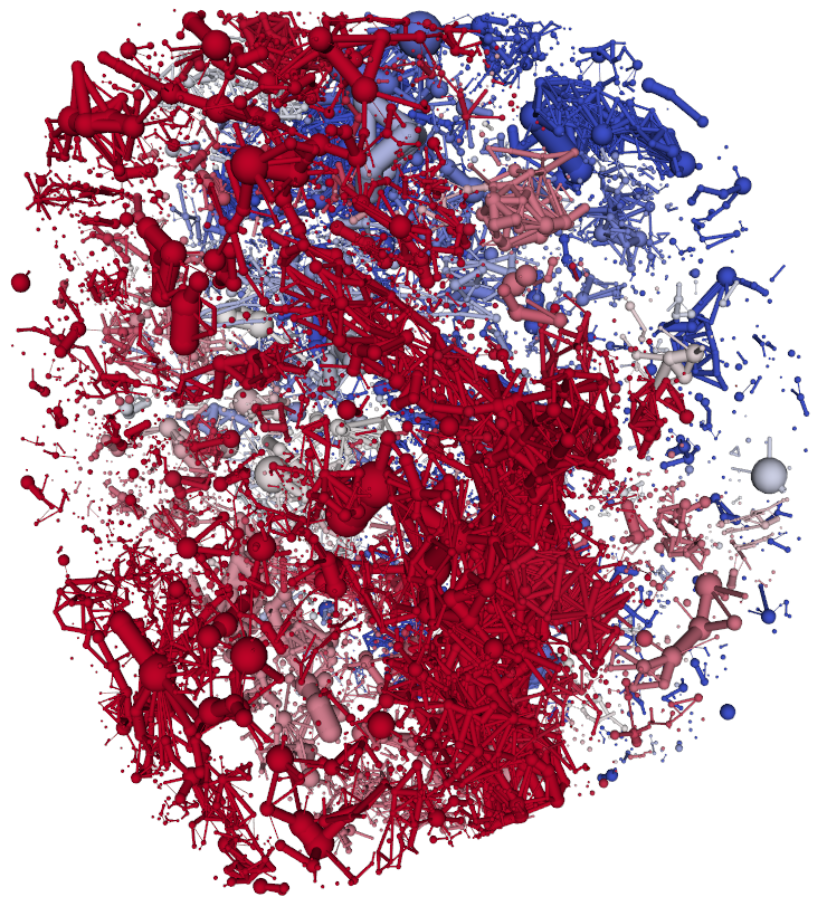}\\[-0.5ex]\footnotesize (a) XPM macropores};

\node[anchor=south west, align=center, inner sep=0pt, outer sep=0pt] (tr) at (0.35\textwidth, 0) {
  \includegraphics[width=0.33\textwidth]{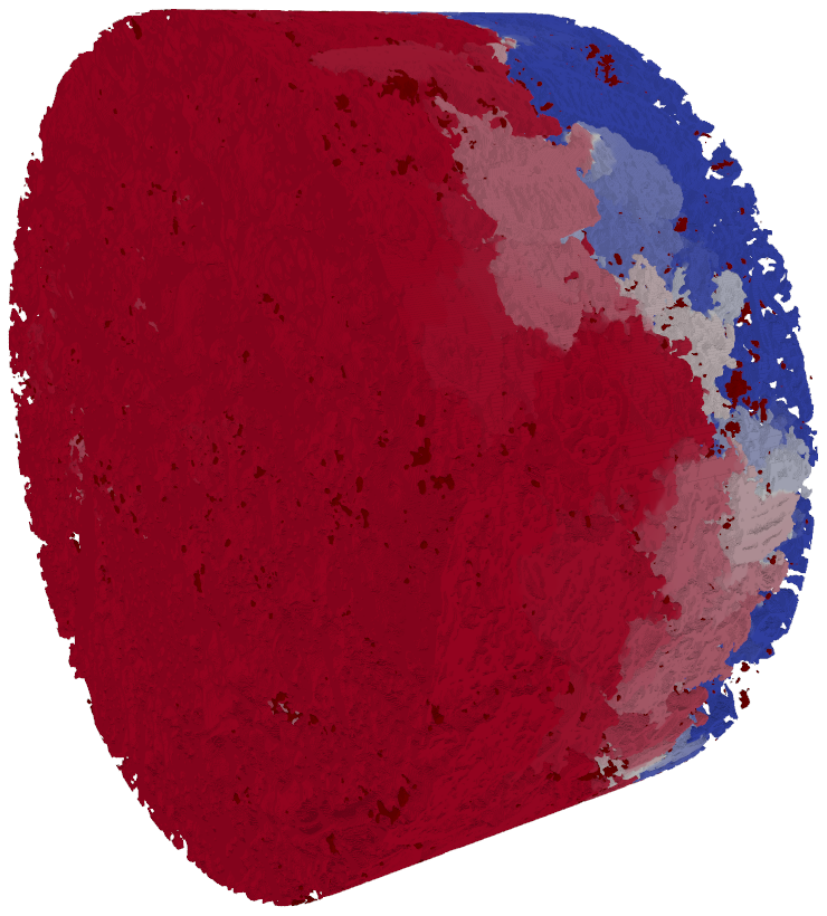}\\[-0.5ex]\footnotesize (b) XPM micropores};

\node[anchor=north west, align=center, inner sep=0pt, outer sep=0pt] at (0, -1em) {
  \includegraphics[width=0.33\textwidth]{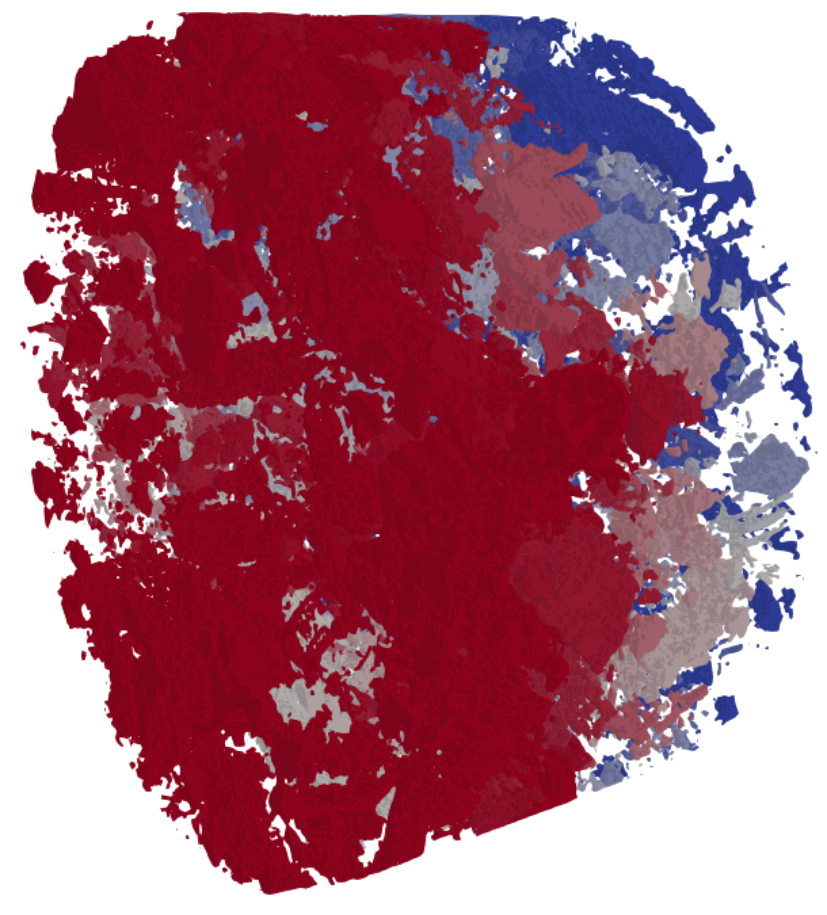}\\[-0.5ex]\footnotesize (c) DNS macropores};

\node[anchor=north west, align=center, inner sep=0pt, outer sep=0pt] at (0.35\textwidth, -1em) {
  \includegraphics[width=0.33\textwidth]{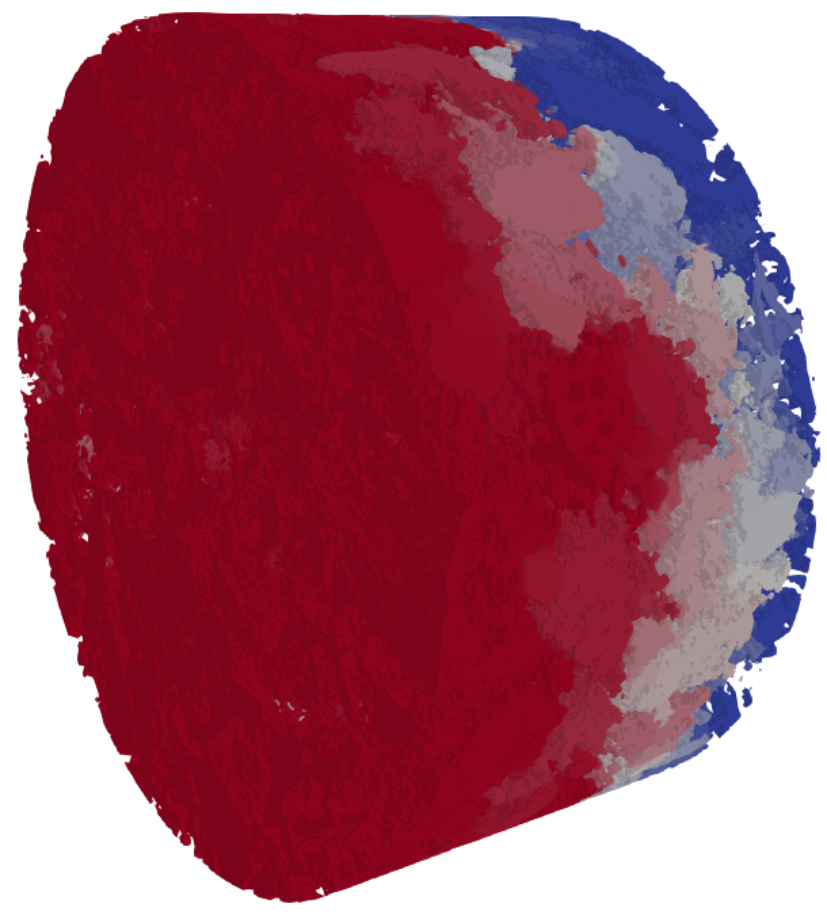}\\[-0.5ex]\footnotesize (d) DNS micropores};

\pressurelegendbar
\begin{scope}[shift={(tr.north west)}]
  \draw[<->] (0.053\textwidth, 0.005\textwidth) -- (0.218\textwidth, 0.005\textwidth) node[midway, above] {\footnotesize 2.75 mm};
\end{scope}
\end{tikzpicture}
\caption{Pressure fields calculated with XPM and  DNS for the MCR-1 sample.}
\label{fig:M1}
\end{figure*}

\begin{figure*}[!t]
\centering
\begin{tikzpicture}[color=black]
\node[anchor=south west, align=center, inner sep=0pt, outer sep=0pt] at (0, 0) {
  \includegraphics[width=0.33\textwidth]{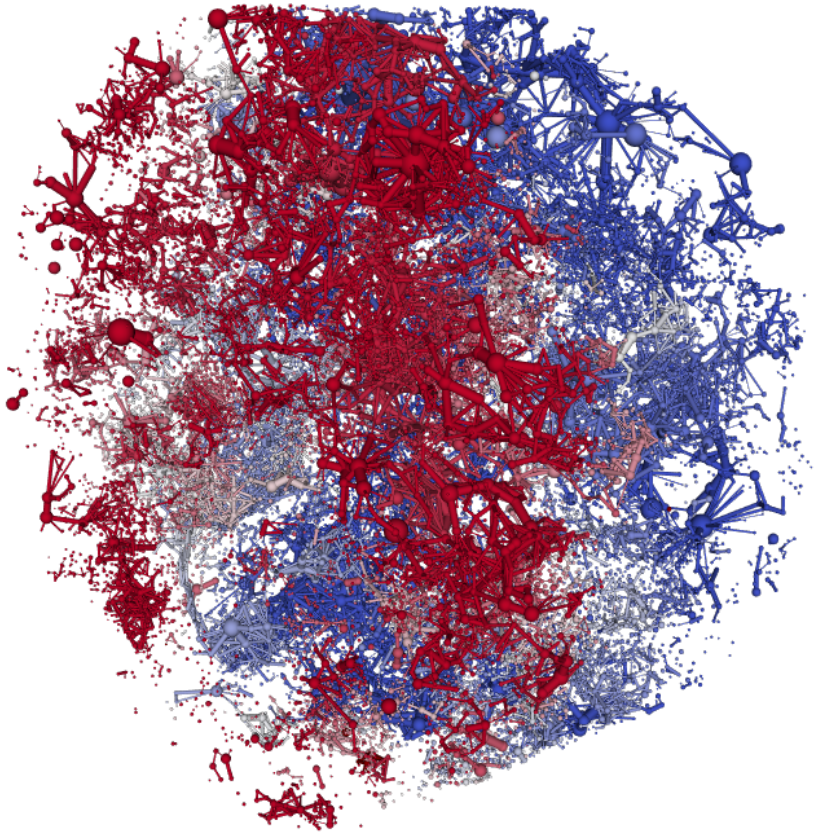}\\[-0.5ex]\footnotesize (a) XPM macropores};

\node[anchor=south west, align=center, inner sep=0pt, outer sep=0pt] (tr) at (0.35\textwidth, 0) {
  \includegraphics[width=0.33\textwidth]{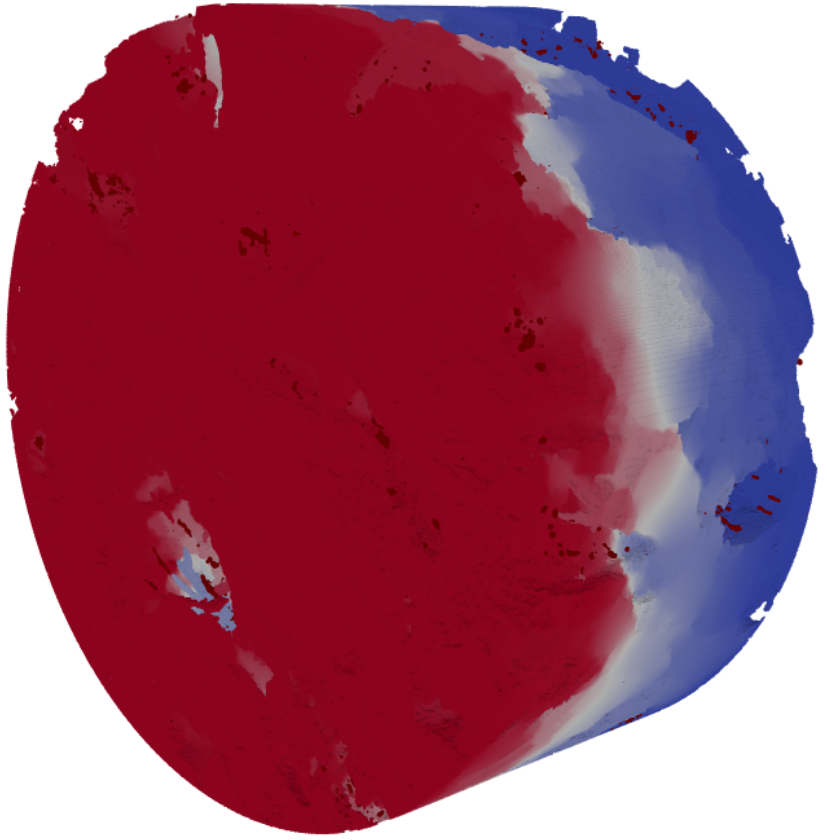}\\[-0.5ex]\footnotesize (b) XPM micropores};

\node[anchor=north west, align=center, inner sep=0pt, outer sep=0pt] at (0, -1em) {
  \includegraphics[width=0.33\textwidth]{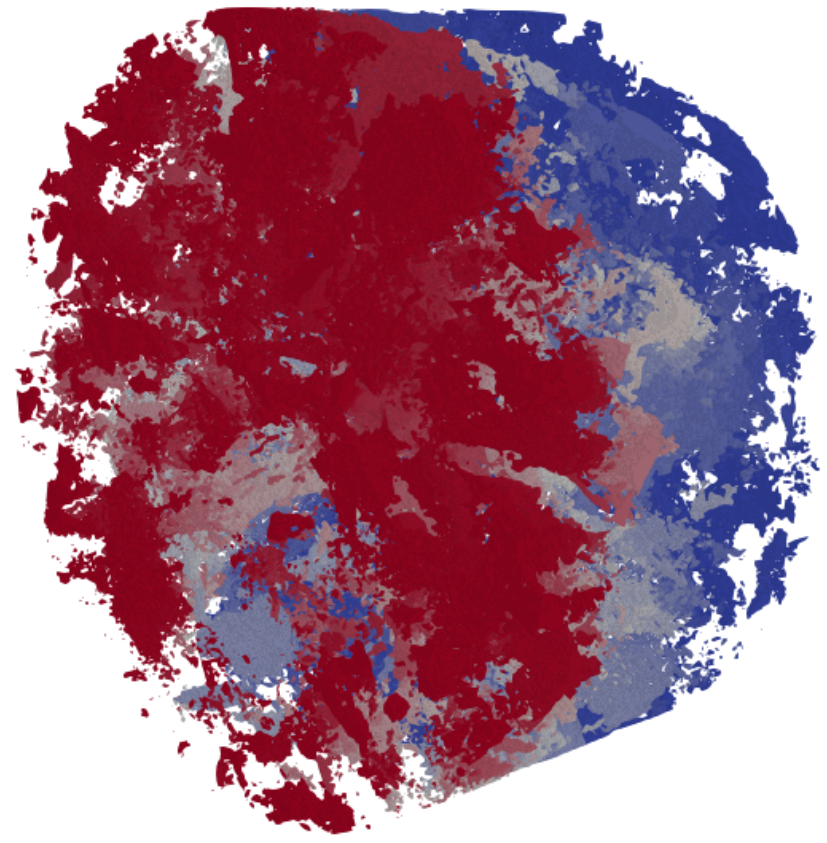}\\[-0.5ex]\footnotesize (c) DNS macropores};

\node[anchor=north west, align=center, inner sep=0pt, outer sep=0pt] at (0.35\textwidth, -1em) {
  \includegraphics[width=0.33\textwidth]{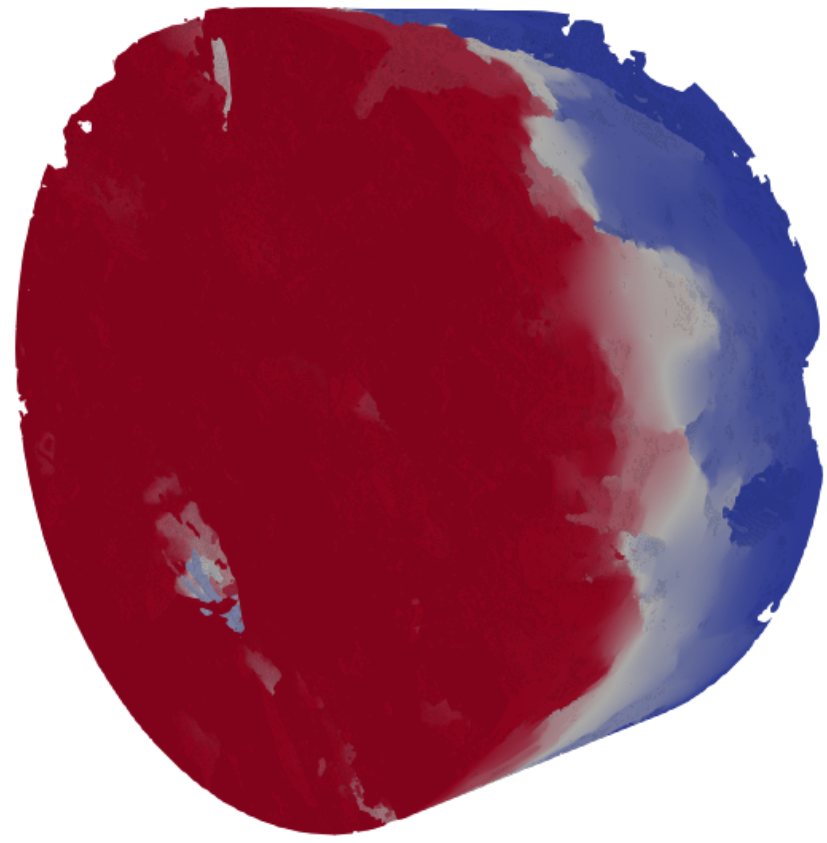}\\[-0.5ex]\footnotesize (d) DNS micropores};

\pressurelegendbar
\begin{scope}[shift={(tr.north west)}]
  \draw[<->] (0.078\textwidth, 0.007\textwidth) -- (0.223\textwidth, 0.007\textwidth) node[midway, above] {\footnotesize 1.98 mm};
\end{scope}
\end{tikzpicture}
\caption{Pressure fields calculated with XPM and DNS for the Estaillades-1 sample.}
\label{fig:EST1}
\end{figure*}

In addition, the obtained pressure fields using \abr{XPM} and \abr{DNS} are compared for \abr{MCR-1} (Figure \ref{fig:M1}) and \mbox{Estaillades-1} (Figure \ref{fig:EST1}). For better visualisation, the macropores and the micropores have been separated. For \abr{XPM}, the macroscopic network is shown on the left, and the Darcy cells are shown on the right. For the \abr{DNS} simulation, the voxels within the macropores ($\varepsilon=1$) are shown on the left, and the voxels within the micropores ($0<\varepsilon<1$) are shown on the right. The macropore images for \abr{DNS} appear denser than those for \abr{XPM}, due to the network being rendered as spheres and cylinders. However, there is an exact voxel correspondence between the two.

We observe a strong correlation between the pressure fields. The fields are more segregated for \abr{MCR-1} and more diffuse for Estaillades-1, which we attribute to the larger number of labels in Estaillades. For \abr{MCR-1}, the pressure field is slightly more diffuse with \abr{DNS}, while the \abr{XPM} simulation shows marginally higher pressure in the bottom right corner of the domain. For Estaillades, the pressure fields are almost identical. 

We conclude that our \abr{PNM} workflow is able to reproduce the results of \abr{DNS} simulations based on the micro-continuum approach with great accuracy, as long as the conductance in the macroscopic network has been evaluated accurately. After correcting the conductance to match the permeability obtained with \abr{DNS} for the macropores, we obtain a good match for total permeability and for pressure fields.

\subsection{Comparison with experiment}

In this section, our \abr{XPM} modelling workflow is compared with the results of permeability experiments on the \abr{MCR} and Estaillades samples, imaged with properties presented in Table \ref{Tab:CT}. For \abr{MCR}, the permeability was obtained by conducting single-phase flow experiments using deionised water with a differential pressure transducer. For Estaillades, the permeability measurement is presented in \citep{Menke2022}. 

\begin{figure*}[!t]
\centering
\begin{tikzpicture}[color=black]
\node[anchor=south west, align=center, inner sep=0pt, outer sep=0pt] (tl) at (0,              0) {
  \includegraphics[height=0.26\textwidth]{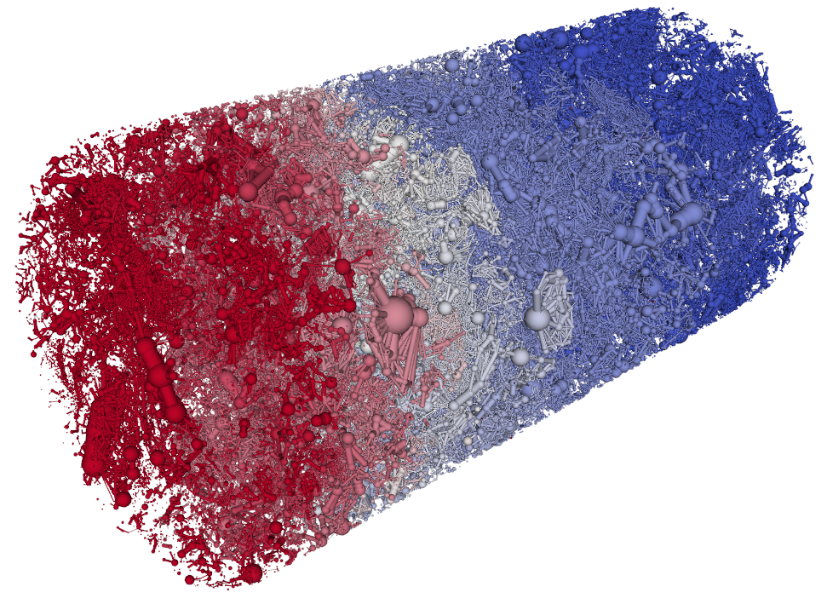}\\[-0.5ex]\footnotesize (a) MCR macropores};

\node[anchor=south west, align=center, inner sep=0pt, outer sep=0pt] (tr) at (0.37\textwidth, 0) {
  \includegraphics[height=0.26\textwidth]{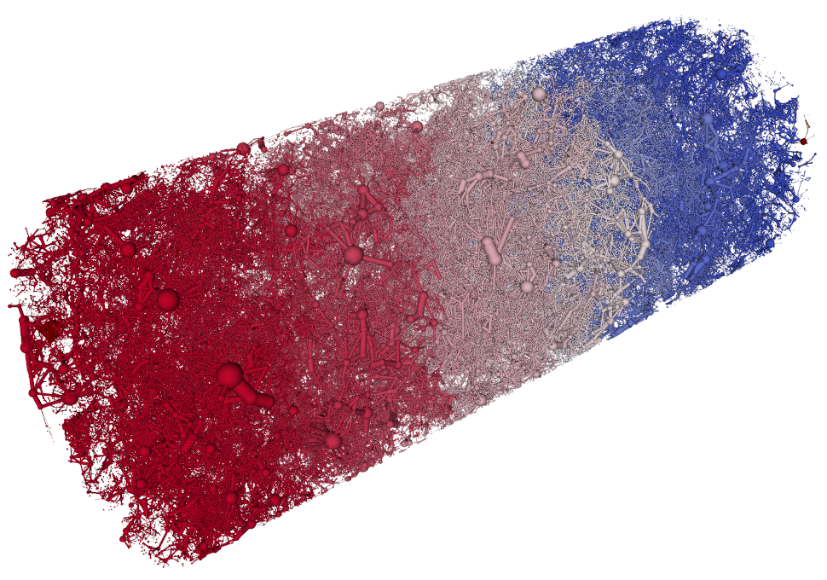}\\[-0.5ex]\footnotesize (b) Estaillades left macropores};

\node[anchor=north west, align=center, inner sep=0pt, outer sep=0pt] at (0,              -1em) {
  \includegraphics[height=0.26\textwidth]{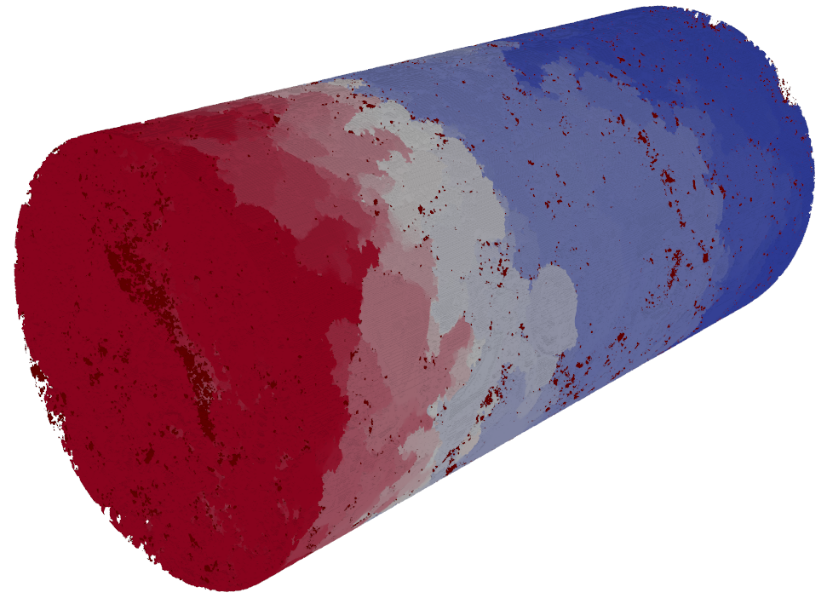}\\[-0.5ex]\footnotesize (c) MCR micropores};

\node[anchor=north west, align=center, inner sep=0pt, outer sep=0pt] at (0.37\textwidth, -1em) {
  \includegraphics[height=0.26\textwidth]{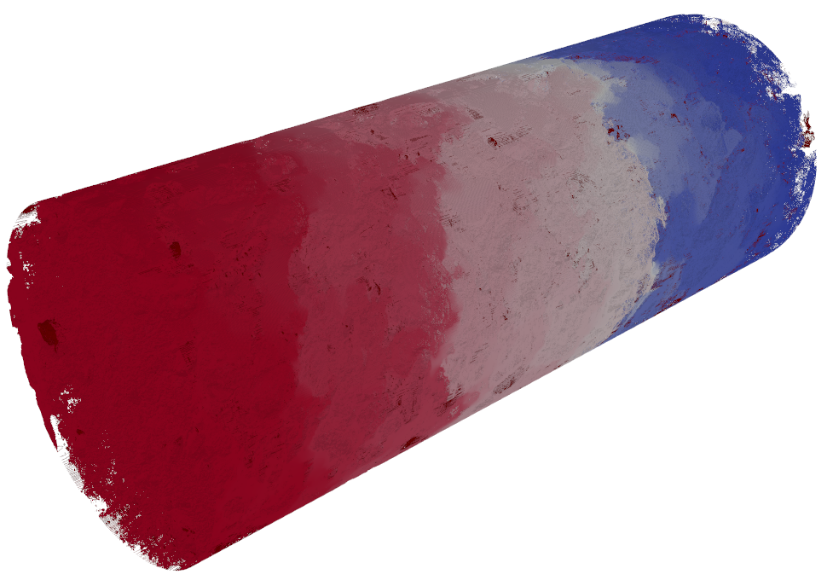}\\[-0.5ex]\footnotesize (d) Estaillades left micropores};

\pressurelegendbar
\begin{scope}[shift={(tl.north west)}]
  \draw[<->] (0.023\textwidth, -0.054\textwidth) -- (0.281\textwidth, 0.007\textwidth) node[midway, above, sloped] {\footnotesize 16.5 mm};
\end{scope}
\begin{scope}[shift={(tr.north west)}]
  \draw[<->] (0.018\textwidth, -0.082\textwidth) -- (0.301\textwidth, 0.001\textwidth) node[midway, above, sloped] {\footnotesize 11.9 mm};
\end{scope}
\end{tikzpicture}
\caption{Pressure fields calculated with XPM for the MCR and Estaillades left samples.}
\label{fig:EstExp}
\end{figure*}

The numerical simulation are performed on a 64-thread Intel Xeon Gold 6448Y Processor with 1 terabyte of memory. With this computing architecture, it was possible to simulate the full \abr{MCR} sample, but the Estaillades sample had to be cut in two, i.e. Estaillades left and right, which are both 3000\texttimes1202\texttimes1218 voxels. Each simulation was done in approximately two hours. The pressure fields for \abr{MCR} and Estaillades left are presented in Figure~\ref{fig:EstExp}.

Table \ref{Tab:ResExp} shows the comparison between the \abr{XPM} permeability and the experimental values. The permeability for the full Estaillades sample has been calculated by harmonic averaged of the left and right values. We observe that \abr{XPM} was able to reproduce the experimental value with great accuracy.

\begin{table}
\caption{Comparison of XPM permeability with experimental values for MCR and Estaillades. The permeability for the full Estaillades sample has been calculated by harmonic averaged of the left and right values.  \label{Tab:ResExp}}
\renewcommand{\arraystretch}{1.5}
\centering
\small
\begin{tabular}{cclc}
\hline
Sample & $k_\text{macro}$ & \multicolumn{1}{c}{$k$} & $k_\text{experiment}$ \\
\hline
MCR & 5.0 & 17.6 & 16 \\
\multicolumn{1}{l}{Estaillades left} & 2.7 & 19.5 & -- \\
\multicolumn{1}{l}{Estaillades right} & 12 & 29.9 & -- \\
\multicolumn{1}{l}{Estaillades full} & -- & 23.6$^\text*$ & 24.6 \\
\hline
\end{tabular}
\vspace{1ex}

$^\text*$Harmonic average between $k_\text{left}$ and $k_\text{right}$ values. Permeabilities are reported in mD.
\end{table}